\documentclass[aps,preprint]{revtex4}
\usepackage{graphicx}

\begin{document}

\title{Generating correlated networks from uncorrelated ones}

\author{A. Ramezanpour}
\email{ramzanpour@sharif.edu}
 \affiliation{Department of Physics, Sharif University of
Technology, P.O.Box 11365-9161, Tehran, Iran.}

\author{V. Karimipour}
 \email{vahid@sharif.edu}
\affiliation{Department of Physics, Sharif University of
Technology, P.O.Box 11365-9161, Tehran, Iran.}

\author{A. Mashaghi}
\email{mashaghi@ibb.ut.edu}
 \affiliation{Institute of Biophysics
and Biochemistry, P.O.Box 13145-1384, Tehran, Iran.}

\date{\today}

\begin{abstract}
In this paper we consider a transformation which converts
uncorrelated networks to correlated ones( here by correlation we
mean that coordination numbers of two neighbors are not
independent). We show that this transformation, which converts
edges to nodes and connects them according to a deterministic
rule, nearly preserves the degree distribution of the network and
significantly increases the clustering coefficient. This
transformation also enables us to relate percolation properties
of the two networks.
\end{abstract}

\maketitle

\section{Introduction}\label{1}
One of the oldest and best studied models of networks which have
the merit of being exactly solvable for many of their properties,
are the random graphs of Erd\"{o}s and R\'{e}nyi \cite{er,b}.
These graphs consist of $N$ nodes any two of which are connected
with a probability $p$ and left unconnected with a probability
$1-p$. Many of the properties of these networks can be easily
derived by analytical means. Among them are the degree
distribution of nodes, which turns out to be Poissonian, the
average shortest path between any two nodes which is of the order
of $\log(N)$, and the onset of phase transition for developing a
giant
component which happens as $p$ exceeds a certain critical value.\\
Despite their exact solvability and their low diameter, these
networks lack some of the other crucial properties of real life
networks. In particular it is well known that many real networks
e.g. World Wide Web, social networks, power grids, scientific and
artistic collaborations, neural and metabolic networks, show
clustering or transitivity which is absent in Erd\"{o}s and
R\'{e}nyi graphs. Moreover many real networks do not possess
Poissonian degree distribution and intensive studies have been
made to construct models as close as to real networks
\cite{ab,s,ws,dm} and to study dynamic effects, e.g. spreading of
a contact effect,
on them.\cite{ka,pv}.\\
In the past few years an elegant theory has been developed to
construct random graphs with desirable degree distributions to
mimic the degree properties of real networks. It appears that
Bender and Canfield \cite{bc,nsw} have been the first to propose
an algorithm for constructing a random graph with a specific
degree distribution.  We will call the ensemble of graphs
constructed in this way the Bender Canfield ensemble. It is
remarkable that these graphs are still exactly solvable to a large
extent \cite{nsw}. However these graphs still have two
shortcomings. First they do not show correlation in the degree of
nearest neighbors and second their clustering
coefficient vanishes in the large $N$ limit.\\
It is important that the degree distribution does not determine by
itself the existence or lack of correlations. For a specific
degree distribution one may have or have not correlations.
Moreover, it has been observed that correlation is an essential
feature of real networks which can appear in different forms
\cite{n}. For instance, a high degree node may be connected to
other high degree nodes(associative mixing), to low degree
nodes(dissociative mixing) or with equal probability to both types
(neutral mixing) with different
resulting behaviors in networks \cite{bp,mnl,kr,pvv}.\\
For this reason algorithms have been developed to produce
correlated networks with certain degree distribution
 \cite{dms1,bl,n,dm}. \\

In this paper we will suggest a simple deterministic
transformation on the Bender Canfield (BC) graphs and show that
the transformed graphs are both correlated and clustered in the
large $N$ limit. Given a BC graph $G$ with $N$ nodes and a degree
distribution $P(k)$, we construct a graph $\tilde{G}$ by assigning
nodes to each edge of $G$. We then connect these new nodes if the
corresponding edges in $G$ have had a common node in $G$. We show
that many of the properties of these transformed networks can be
obtained exactly or almost exactly. We obtain general formulas
for the degree distribution and its correlations for the
transformed graphs and will obtain also formulas for the
clustering coefficients of these new graphs. As examples we apply
our transformation to Bender Canfield graphs with various degree
distributions.\\
It has been shown by Newman that percolation on BC graphs can be
solved by a generating function method. The method is applicable
due to the fact that in these graphs there is no clustering. By
applying our transformation to these graphs we can follow similar
steps and solve percolation on $\tilde{G}$ . The interesting point
is that now $\tilde{G}$ is a highly clustered graph for
which we can solve percolation.\\
The paper is structured as follows: In section \ref{2} we give a
brief review of Bender Canfield ensemble of random graphs having
arbitrary degree distributions. In section \ref{3} we discuss our
transformation and derive various properties of general
transformed graphs. In section \ref{4} we apply our general
formulas to graphs with degree distributions of Poissonian, scale
free and exponential types. We end the paper with a conclusion.

\section{Bender-Canfield ensemble of graphs}\label{2} Let $G$
denote a graph with $N$ edges and $L$ links. Let also the degree
distribution of this graph be given by the function $P(k)$, that
is the fraction of nodes with $k$ neighbors be given by $ P(k)$.
There is an algorithm \cite{bc,nsw} for constructing graphs whose
degree distribution corresponds to $P(k)$ for large $N$. More
specifically given a degree sequence $(k_1, k_2, \cdots k_N)$
corresponding to the desired degree distribution $P(k)$, one takes
each node $i$ with $k_i$ loose ends (stubs) and then connects each
pair of stubs randomly until no loose end remains. We call these
types of graphs Bender-Canfield or simply BC graphs. Thus we speak
of Poissonian BC graphs or scale free BC graphs to designate
the degree distribution used for their construction.\\
Many of the properties of BC graphs can be calculated exactly.
For example the average number of first neighbors of an arbitrary
node, denoted by $z_1$ is given by
\begin{equation}\label{z1}
  z_1 := \langle k \rangle = \sum_k k P(k).
\end{equation}
It is useful to call a node with $k$ emanating edges a node of
type $k$. Then the probability of picking up a node of type $k$
is given by $P(k)$. We can now ask a different question: What is
the probability $q(k)$ of picking up an {\it{edge}} which belongs
to a node of type $k+1$. This is equal to the fraction of stubs
coming out of nodes of type $k+1$:
\begin{equation}\label{q(k)}
  q(k) = \frac{(k+1)P(k+1)}{\sum_k k P(k)}\equiv \frac{(k+1)P(k+1)}{\langle k \rangle}
\end{equation}
If we now follow a link to one of its ends the average number of
new links ( the average ratio of the number of second to the first
neighbors of an arbitrary node $\frac{z_2}{z_1}$) will be given by
\begin{equation}\label{z2/z1}
\frac{z_2}{z_1} = \sum_k kq(k) =\frac{\langle k^2 \rangle -
\langle k \rangle.}{\langle k\rangle}
\end{equation}
As we will see this quantity will play a central role in many of
the later derivations.\\

We can ask yet another question. What is the probability $P(k,k')$
of picking up an edge which is common to a node of type $k+1$ and
a node of type $k'+1$?. This probability is given by a product
which is a reflection of the absence of correlations in these
networks,
\begin{equation}\label{p(k,k')}
 P(k,k')=q(k)q(k').
\end{equation}
One can also calculate the clustering coefficient of these
graphs. The result is \cite{nsw}
\begin{equation}\label{C}
  C = \frac{z_1}{N}(\frac{\langle k^2\rangle- \langle k \rangle}{\langle k
  \rangle^2})^2 = \frac{1}{n}\frac{(z_2)^2}{(z_1)^3}.
\end{equation}
It is important to note that for many kinds of degree
distributions (i.e. those with finite $z_2$ and $ z_1$), this
clustering coefficient vanishes in the limit of large graph size
$N$.  It can also be shown that the ratio $\frac{z_2}{z_1}$
controls the existence of an infinite cluster of connected nodes
\cite{nsw,dms2} for these graphs. For $\frac{z_2}{z_1}>1$ there
is an infinite cluster where the average distance between two
arbitrary nodes, is of order $log(n)$. Recently it has been found
that almost all pairs of nodes have the same distance in this
cluster \cite{dms2}.\\ It will be convenient to define two
generating functions for the BC ensemble corresponding to a
degree distribution:
\begin{equation}\label{gf1}
  G_0(t):= \sum_{k}t^kP(k) \hskip 2cm  G_1(t):= \sum_{k}t^kq(k) =
  \frac{1}{z_1}G_0'(t)
\end{equation}
where $G'(t) = \frac{d}{dt} G(t)$ and in the last relation we have
used (\ref{q(k)}). In terms of these generating functions, the
average number of first neighbors $z_1$ and second neighbors $z_2$
are simply given by
\begin{equation}\label{z1z2}
  z_1 = \frac{d}{dt}G_0(t)|_{t=1}\hskip 2cm  z_2 =
  z_1 \frac{d}{dt}G_1(t)|_{t=1}= G_0''(1).
\end{equation}
\section{Transformation of BC graphs}\label{3} 
\begin{figure}
  \centering
  \includegraphics[width=17cm,height=10cm]{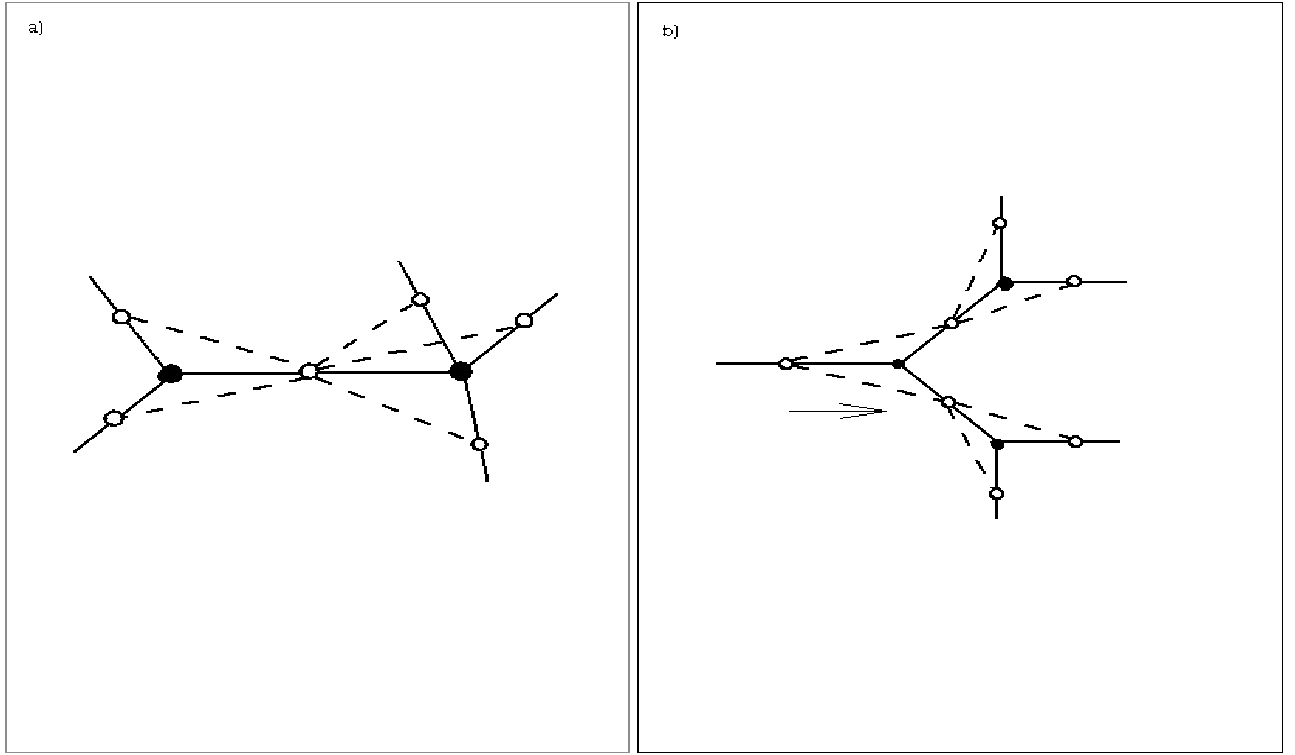}
    \caption{a) First neighbors of a node in $\tilde{G}$.
    b) Second neighbors of a node in $\tilde{G}$( reachable from
    one of its sides).}\label{fig-2}
    \end{figure}

Consider a BC
graph $G$ with $N$ nodes with a degree sequence $(k_1, k_2, \cdots
k_N)$ taken from a distribution $ P(k)$. We can transform this
graph to a new graph $ \tilde{G}$ as follows. To each link of the
original graph, we assign a node of the new graph. We connect any
two nodes of the new graph if the corresponding links in the
original graph have a node in common, see fig.(\ref{fig-1}). The
number of nodes and links of $\tilde{G}$ denoted respectively by $
\tilde{N}$ and $\tilde{L}$ respectively are determined from the
degree distribution of $G$. We note that each node of type $k$ of
$G$ contributes $k$ nodes and $ k(k-1)/2$ edges to $\tilde{G}$.
Taking care of the fact that the new nodes are counted twice we
find:

\begin{figure}
  \centering
  \includegraphics[width=13cm,height=10cm]{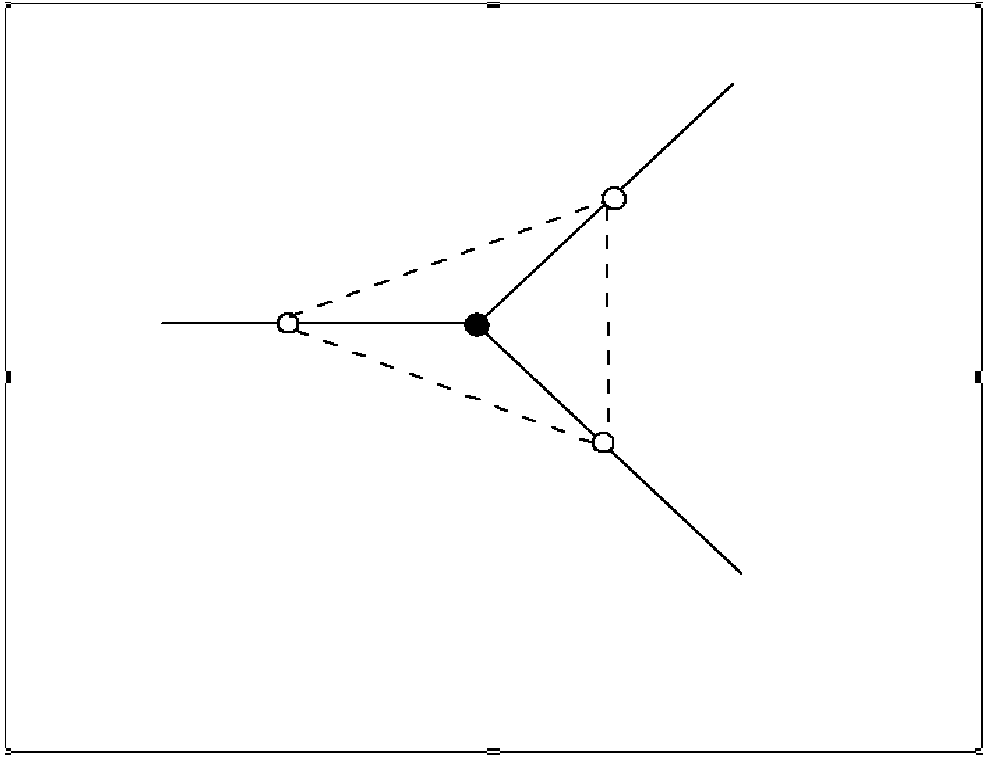}
    \caption{The basic transformation. The filled circle and solid lines
     belong to $G$ and empty circles and dashed lines belong to $\tilde{G}$.}\label{fig-1}
\end{figure}

\begin{eqnarray}\label{tildeN}
  \tilde{N} &=& \frac{1}{2} \sum_{i=1}^{N}k_i \\ \nonumber
  \tilde{L} &=& \frac{1}{2} \sum_{i=1}^N k_i(k_i-1)
\end{eqnarray}
As figure (\ref{fig-2}-a) shows the degree distribution of the new
graph $ \tilde{G}$ is given by:
\begin{equation}\label{pk}
    \tilde{P}(k)=\sum_{r+s=k}q(r)q(s),
\end{equation}
which is nothing but the probability that an arbitrary edge in
$G$ is connected to a total of $k$ edges at its two end point nodes.\\
There are simple relations between the generating functions of $G$
and $\tilde{G}$. Using (\ref{pk}) we find:

\begin{equation}\label{gf2}
  \tilde{G_0}(t) = \sum_k t^k \tilde{P}(k) = \sum_k t^{k}
  \sum_{r+s=k} q(r)q(s) = \sum_{r,s} t^{r+s}
  q(r)q(s)= (\frac{G_0'(t)}{z_1})^2
  \end{equation}

From this last equation we find
\begin{equation}\label{tildez1}
  \tilde{z_1}:= \frac{d}{dt}\tilde{G_0}(t)|_{t=1} = 2 \frac{\langle k^2 - k\rangle}{\langle k
  \rangle}= 2 \frac{z_2}{z_1}
\end{equation}
In view of (\ref{gf1}) we  have
\begin{equation}\label{tildeq}
  \tilde{G_1}(t) :=\frac{1}{\tilde{z_1}}\frac{d}{dt}\tilde{G_0}(t)= \frac{G_0'(t)G_0''(t)}{z_1
  z_2}.
\end{equation}

\begin{figure}
\centering
  \includegraphics[width=15cm,height=10cm]{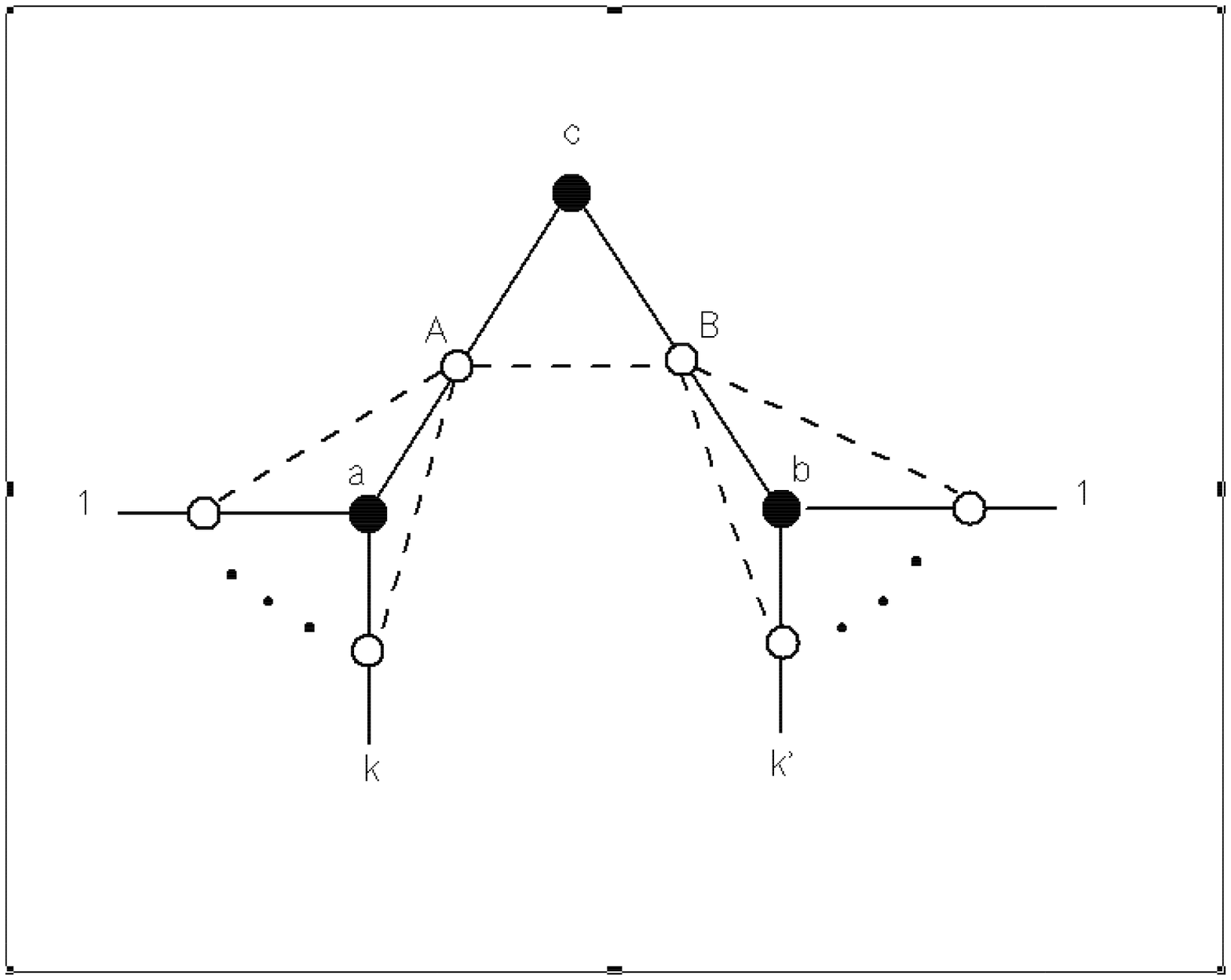}
    \caption{The degrees of the nodes $A$ and $B$ in $\tilde{G}$
    are correlated by the vertex $c$ in $G$.}
    \label{fig-3}
\end{figure}

However since the graph $\tilde{G}$ is a clustered graph as we
will see, the ratio of average number of second to first neighbors
of an arbitrary node, $ \frac{\tilde{z_2}}{\tilde{z_1}} $ is not
given by the expression $ \sum_k k \tilde{q}(k)$ as in
(\ref{z2/z1}).  Instead we resort to a direct counting. As shown
in fig. (\ref{fig-2}-b) if we follow a node of $\tilde{G}$ to the
right, the number of first neighbors that we find is given by
$\frac{z_2}{z_1}$ inherited from $G$. Due to the low clustering of
$G$, the number of second neighbors that we will meet will be
$(\frac{z_2}{z_1})^2 $. Thus the total number of second neighbors
will be twice this value, that is $\tilde{z_2} =
2(\frac{z_2}{z_1})^2 $. The above reasoning indeed shows that
$\frac{\tilde{z_2}}{\tilde{z_1}} = \frac{z_2}{z_1} $ which in turn
means that the conditions \cite{nsw} for the development of a
giant component in $G$ and$ \tilde{G}$ are identical.\\
We now find the probability $\tilde{P}(k,k')$  of finding an edge
the end nodes of which have $k$ and $k'$ other neighbors.\\
Looking at figure (\ref{fig-3}), the probability of finding an
edge like $AB$ in $\tilde{G}$ is equivalent to finding two edges
$ac$ and $bc$ incident on the same node $c$ in $G$. For a moment
suppose that no other edge in $G$ is incident on $c$. Then it is
clear from figure (\ref{fig-3}) that the nodes $A$ and $B$ will
have $k+1$ and $k'+1$ neighbors in $\tilde{G}$ if the nodes $a$
and $b$ will have $k+1$ and $k'+1$ neighbors in $G$ respectively.
Putting this together we find that in this simple case
$\tilde{P}(k,k') = P(2)q(k)q(k')$, where $P(2)$  comes from the
probability of finding a node like $c$ of degree $2$ in $G$. In
general the node $c$ may be common to $t$ other edges in $G$.
These extra edges contribute to the total number of neighbors of
$A$ and $B$, so that in order for the node $A$ to have a total of
$k+1$ neighbors in $\tilde{G}$, the node $a$ needs only have
$k+1-t$ neighbors in $G$. A similar statement is true also for the
node $B$. Thus instead of the factor $q(k)q(k')$ we will have $
q(k-t)q(k'-t)$. This should be multiplied by the probability of
finding a triplet $acb$ which is proportional to $
\frac{(t+2)(t+1)}{2} P(t+2)$ and finally summed over $t$. The
final result is
\begin{equation}\label{pkk}
\tilde{P}(k,k')=\sum_{t}
\frac{(t+2)(t+1)}{z_2}P(t+2)q(k-t)q(k'-t),
\end{equation}
In this way correlations are introduced into the graph in the
sense that $P(k,k')$ is no longer equal to $ q(k)q(k')$.\\
It is also possible to calculate the clustering coefficient of
$\tilde{G}$. It is clear that an edge with end point nodes of
degree $r+1$ and $s+1$ in $G$ represents a node of degree $r+s$ in
$\tilde{G}$. Thus the total number of potential connections among
these first neighbors is $\frac{(r+s)(r+s-1)}{2}$. Of these
possible connections , there are already  a number of
$\frac{r(r-1)}{2}+\frac{s(s-1)}{2}$ connections present coming
from the definition of $\tilde{G}$. Due to the clustering
coefficient of $G$, there are configurations which increase this
number for finite graphs. However since in the thermodynamic limit
we know that the BC graphs have vanishing clustering coefficients
we need not worry about these contributions. Thus in the
thermodynamic limit we have the following formula for clustering
coefficient of $\tilde{G}$:

\begin{equation}\label{c}
    C=\sum_{r,s}\frac{r(r-1)+s(s-1)}{(r+s)(r+s-1)}q(r)q(s).
\end{equation}
In this way our transformation has introduced a finite clustering
coefficient into the BC ensemble of graphs. In the next section we
will apply this transformation to several well known ensembles
with specific degree distributions, namely the
Poisson, scale free and exponential ensembles.\\

Finally let us consider percolation on $\tilde{G}$. For the sake
of simplicity, here we consider only site percolation but the same
analysis can be applied to bond percolation as well. Let each node
of $\tilde{G} $ be occupied with probability $p$ and denote the
probability that an arbitrary node belongs to a cluster of size
$n$, by $P_n$. The generating function of this probability is
denoted by $ H(x) $, that is $ H(x) = \sum_{n=0}^\infty P_n x^n$.
Using the same procedure as in \cite{cnsw} we write the following
expression for $H(x)$ :
\begin{equation}\label{H}
    H(x)=1-p+pxh^2(x),
\end{equation}
where $h(x)$ is the generating function for the number of nodes
reachable if we follow the neighbors of the node in one of its
sides say to the right (i.e. if we follow the corresponding link
on $G$ to the right) (see fig. (\ref{fig-2}-b)). The expression
for $ h(x)$ is obtained recursively as
\begin{eqnarray}\label{z}
  h(x) &=& \sum_{k=0}^\infty q(k) \sum_{r=0}^k (\begin{array}{c}
    k \\
    r \
  \end{array}) p^r (1-p)^{k-r}h^r(x)x^r \nonumber\\
  &=& G_1(x p h(x) + 1-p)
\end{eqnarray}
Solution of these two equations will give us the probabilities
$P_n$.

\section{Examples}\label{4}
\subsection{Poisson graphs}
\begin{figure}
  \centering
  \includegraphics[width=15cm,height=8cm]{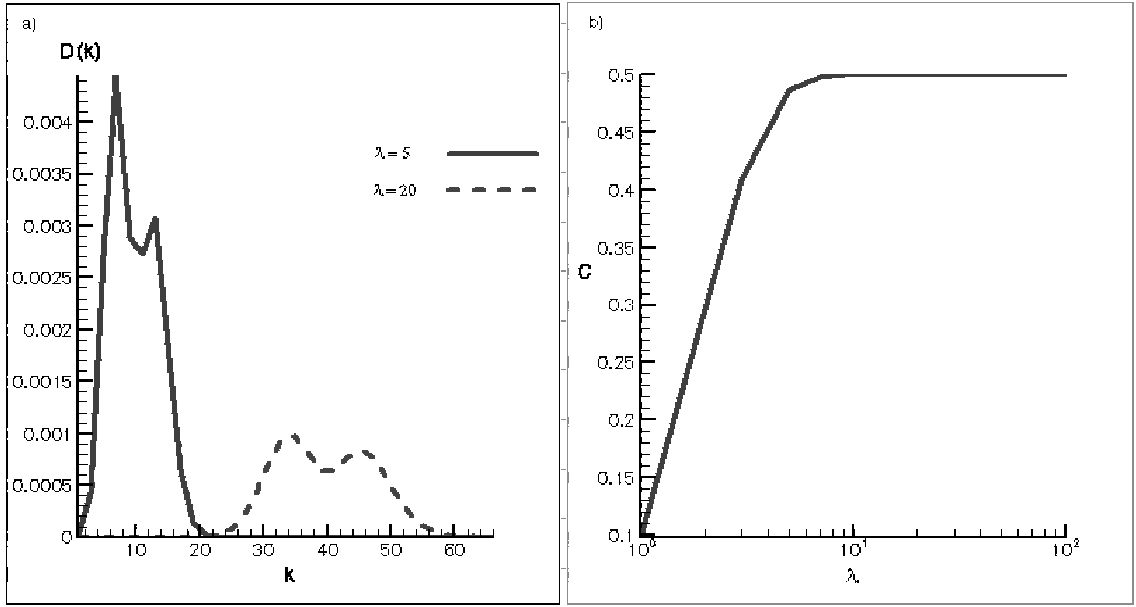}
    \caption{
    a) plots of $D(k)$ for several values of $\lambda$(here
    $k_{max}=1000$) and
    b) Clustering coefficient.}\label{fig-4}
\end{figure}

For a poissonian distribution, where
$P(k)=\frac{\lambda^{k}}{k!}e^{-\lambda}$, it is readily verified
using () that $q(k) = p(k)$. We find from (\ref{pk}) that
\begin{equation}\label{piosson}
\tilde{P}(k) = \sum_{r} P(r)P(k-r) =
\frac{\lambda^{r}}{r!}e^{-\lambda}\frac{\lambda^{(k-r)}}{(k-r)!}e^{-\lambda}=
\frac{(2\lambda)^{k}}{k!}e^{-2\lambda}
\end{equation}
where in the last step we have used the binomial distribution.
Thus our transformation maps a poissonian graph to another
poissonian graph, whose average degree is twice the original one.
However this new graph is completely different from the original
one in other respects. First there is correlations between the
degree of neighbors and second it has a finite clustering
coefficient even for large graphs (as $ N\longrightarrow \infty$).
To see this we use (\ref{pkk}) to calculate for the specific case
of $ k=k'$, the difference $ D(k):=
\tilde{P}(k,k)-\tilde{q}(k)\tilde{q}(k) $ as a function of $k$ for
various values of $ \lambda$. The result is shown in fig.
(\ref{fig-4}-a). Figure (\ref{fig-4}-b) shows the clustering
coefficient (calculated from (\ref{c})) as a function of
$\lambda$. It is clearly seen in this and the other cases
considered below that there are non-vanishing correlations in the
degree distributions. Also these transformed graphs have
appreciable value of clustering. Moreover it is seen that the
clustering coefficient approaches a maximum value of nearly $0.5$
for large value of average connectivity. The reason is that, in
this limit an appreciable fraction of the nodes of $G$ have a high
degree approximately equal to $z_1$,  and thus one can estimate
$C$ from (\ref{c}) as $ C \sim
\frac{2z_1(z_1-1)}{2z_1(2z_1-1)}\sim 0.5 $. This explanation
applies to the other examples discussed below.

\subsection{Scale free graphs}
\begin{figure}
  \centering
  \includegraphics[width=15cm,height=8cm]{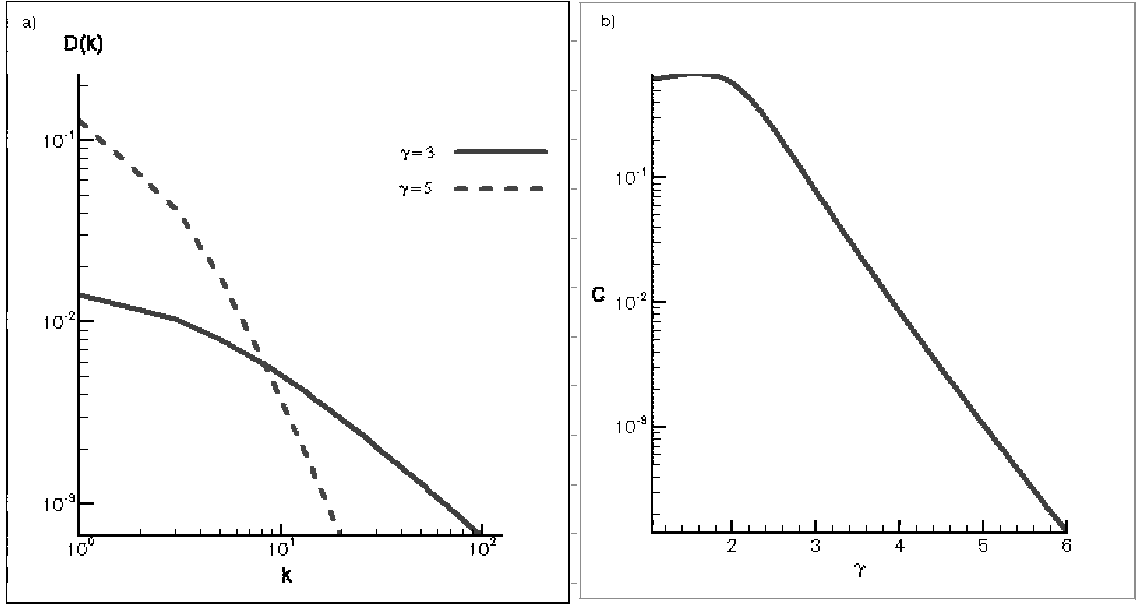}
    \caption{
    a) $D(k)$ and
    b) C for scale free graphs.}\label{fig-5}
\end{figure}

For scale free graphs we have $
P(k)=\frac{1}{\zeta(\gamma)}k^{-\gamma}$ where
$\zeta(\gamma)=\sum_{k=1}^{\infty}k^{-\gamma}$. From
(\ref{z1},\ref{q(k)}) we find $z_1 =
\frac{\zeta(\gamma-1)}{\zeta(\gamma)}$ and $ q(k) =
\frac{(k+1)^{1-\gamma}}{\zeta(\gamma-1)}$. We find from
(\ref{pk}):
\begin{equation}\label{}
  \tilde{P}(k) = \frac{1}{\zeta^2(\gamma-1)}\sum_{s=0}^k
  ((s+1)(k+1-s))^{1-\gamma}
\end{equation}
It is seen that for $k \gg 1$ the above sum is dominated by its
first and last terms. Thus for large $k$,  $\tilde{P}(k) $ behaves
like $ k^{1-\gamma}$ which in turn gives
$\tilde{\gamma}=\gamma-1$. Thus the transformation maintains the
power law behavior of degree distribution for large degrees. To
see this behavior more precisely, we go to the continuum limit
and convert the above sum into an integral which after a little
rearrangement can be cast into the form:
\begin{equation}\label{x}
\tilde{P}(k) = (\gamma-2)^2 \int_{0}^{k} dx
[(\frac{k}{2}+1)^2-(x-\frac{k}{2})^2]^{1-\gamma}
\end{equation}
A change of variable $ x-\frac{k}{2} = (\frac{k}{2}+1)\sin \theta
$ turns this integral into the form
\begin{equation}\label{y}
  \tilde{P}(k) = 2(\gamma -2)^2(\frac{k}{2}+1)^{3-2\gamma}
  \int_{0}^{\alpha}{\cos \theta}^{3-2\gamma} d\theta
\end{equation}
where $ \sin \alpha = \frac{k}{k+2}$. As an example for the case
$ \gamma = 5/2$ we find
\begin{equation}
\tilde{P}(k)= \frac{k}{(k+2)^2\sqrt{k+1}},
\end{equation}
which as  expected, behaves like $k^{-\frac{3}{2}}$ for large $k$.
Like the previous example, we calculated numerically $D(k)$ and
$C$. The results have been shown in figure (\ref{fig-5}).

\subsection{Exponential graphs}
Finally, let us consider exponential distributions, $P(k)= A
e^{-\frac{k}{k_0}}$ where $ A= 1-e^{-\frac{1}{k_0}}$ is a
normalizing factor. We find from (\ref{q(k)}) $  q(k) =
\frac{A^2}{1-A} (k+1)e^{-\frac{k+1}{k_0}}$. Using (\ref{pk}) we
find

\begin{equation}\label{pk3}
\tilde{P}(k) = \frac{A^4}{(1-A)^2} e^{-\frac{k+2}{k_0}}
\sum_{s=0}^{k} (s+1)(k+1-s).
\end{equation}
Converting the sum to integral, we find:
\begin{equation}\label{exp}
\tilde{P(k)}= \frac{A^4}{(1-A)^2} e^{-\frac{k+2}{k_0}}
    (\frac{k^3}{6}+k^2+k).
\end{equation}
We see that transformation does not change the cutoff value,
$k_0$, but produces some polynomial terms.\\

\begin{figure}
  \centering
  \includegraphics[width=15cm,height=8cm]{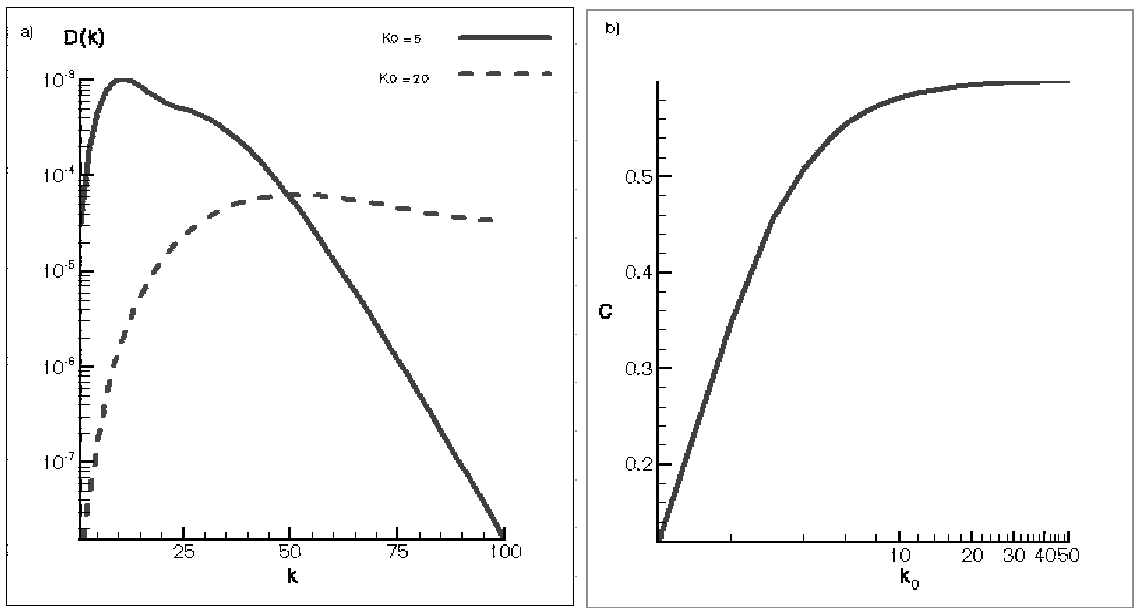}
    \caption{
    a)$D(k)$ and
    b)C for exponential degree distribution.}\label{fig-6}
\end{figure}

In figure (\ref{fig-6}) we display the correlation and the
clustering coefficient for this type of degree distribution, using
equations (\ref{pkk}) and (\ref{c}).

\section{Conclusion}\label{5}
We have introduced a transformation which when applied to an
uncorrelated network with low clustering, produces a new
correlated network with a considerable clustering. The small
world property of the graph is not affected by this
transformation, since the shortest path on two nodes of $G$ is
almost the same as the shortest path on two incident links on
these nodes on $G$ or two nodes on $\tilde{G}$. We thus have a
method to produce ensemble of graphs which resemble more closely
the real networks while still being solvable in many respects.
Moreover it will be possible to use this transformation and solve
dynamic processes on these new graphs. For example we have shown
how site percolation
can be solved on these transformed graphs. \\

This transformation can also be applied to already correlated
networks. Moreover we have considered only the deterministic form
of the transformation . In some cases it may be useful to
introduce a stochastic parameter in the transformation to have
more degree of freedom. That is the nodes of the transformed graph
$\tilde{G}$, can be connected with a probability $p$ if the
corresponding edges in $G$ have a common
node. \\
We hope to investigate these issues in subsequent works.

\end{document}